\documentclass[12pt]{article}

\usepackage{scicite}
\usepackage{times}

\usepackage{graphicx}

\newcommand{\lsim}{{\;\raise0.3ex\hbox{$<$\kern-0.75em\raise-1.1ex\hbox{$\sim$}}\;}}
\def\bra#1{\mathinner{\langle{#1}|}}
\def\ket#1{\mathinner{|{#1}\rangle}}
\def\braket#1{\mathinner{\langle{#1}\rangle}}

\newenvironment{sciabstract}{%
\begin{quote} \bf}
{\end{quote}}

\title{Quantum amplification of mechanical oscillator motion} 

\author
{S. C. Burd$^{1,2,\ast}$, R. Srinivas$^{1,2}$, J. J. Bollinger$^{1}$, A. C. Wilson$^{1}$, D. J. Wineland$^{1,2,3}$, 
\\D. Leibfried$^{1}$, D. H. Slichter$^{1}$, D. T. C. Allcock$^{1,2,3}$
\\
\\
\normalsize{$^{1}$Time and Frequency Division, National Institute of Standards and Technology,}\\
\normalsize{Boulder, CO 80305, USA}\\
\normalsize{$^{2}$Department of Physics, University of Colorado, Boulder, CO 80305, USA}\\
\normalsize{$^{3}$Department of Physics, University of Oregon, Eugene, OR 97403, USA}\\
\\
\normalsize{$^\ast$To whom correspondence should be addressed; E-mail:  shaun.burd@colorado.edu.}
}

\date{}

\begin{document} 

\baselineskip24pt

\maketitle 

\begin{sciabstract}
Detection of the weakest forces in nature and the search for new physics are aided by increasingly sensitive measurements of the motion of mechanical oscillators. However, the attainable knowledge of an oscillator's motion is limited by quantum fluctuations that exist even if the oscillator is in its lowest possible energy state. Here we demonstrate a widely applicable technique for amplifying coherent displacements of a mechanical oscillator with initial magnitudes well below these zero-point fluctuations. When applying two orthogonal ``squeezing" interactions before and after a small displacement, the displacement is amplified, ideally with no added quantum noise. We implement this protocol with a trapped-ion mechanical oscillator and measure an increase of up to 17.5(3)\,decibels in sensitivity to small displacements.
\end{sciabstract}

Mechanical oscillators are essential components in an increasing variety of precision sensing applications including gravitational wave detection~\cite{Abbott2016}, atomic force microscopy~\cite{Butt2005}, cavity optomechanics~\cite{Aspelmeyer2014}, and measurement of weak electric fields~\cite{Ivanov2016,Shaniv2017,Blums2018,Gilmore2017,Wolf2018}. Quantum mechanically, any harmonic oscillator can be described by a pair of non-commuting observables, typically position and momentum for a mechanical oscillator. The precision of measurements of these observables is limited by unavoidable quantum fluctuations that are present even if the oscillator is in its ground state.  Using the method of ``squeezing", these zero-point fluctuations can be manipulated, while preserving their product as dictated by the Heisenberg uncertainty relation. This allows for improved measurement precision for one observable at the expense of increased fluctuations in the other~\cite{Caves1981}. Although squeezed states have been created in a variety of physical systems including optical~\cite{Slusher1985} and microwave fields~\cite{Movshovich1990}, spin systems~\cite{Pezze2018}, micro-mechanical oscillators~\cite{Wollman2015,Lecocq2015,Pirkkalainen2015}, and the motional modes of single trapped ions~\cite{Meekhof1996, Kienzler2015, Maslennikov2017}, exploiting squeezing for enhanced metrology has been challenging. In particular, noise added during the detection process will limit the metrological enhancement unless it is smaller than the squeezed noise. The requirement of low-noise detection can be overcome by increasing the magnitude of the signal to be measured \cite{Hosten2016b}. In optical interferometry~\cite{Yurke1986} and in spin systems~\cite{Davis2016}, it has been shown that reversal of squeezing interactions can magnify small phase shifts, significantly relaxing detection requirements~\cite{Linnemann2016}. Photon field displacements in microwave cavities have also been amplified using similar phase-sensitive amplification schemes~\cite{Malnou2018, Eddins2018}. However, the challenge of implementing reversible squeezing interactions in mechanical oscillator systems has prevented prior use of such methods. 

\begin{figure*}
\centering
\includegraphics[scale=1.0]{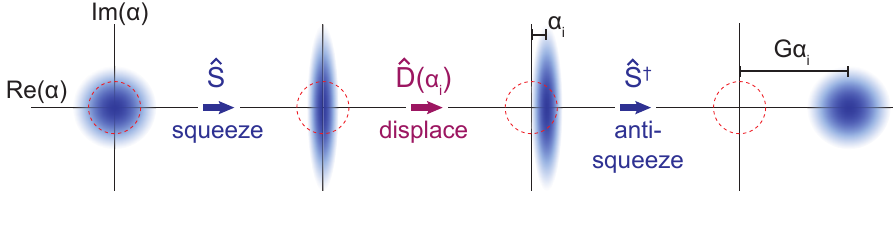}
	\caption{Conceptual illustration of the amplification protocol. Each panel shows a Wigner function phase space distribution (not to scale) in a frame rotating at the oscillator frequency. A displacement $\alpha_i$ of an initially squeezed ground state is amplified by subsequent reversed squeezing (``anti-squeezing"), resulting in a final coherent state with amplitude $G\alpha_i$ with no added noise. Dashed red circles indicate the characteristic extent of the initial ground state fluctuations.}
    \label{fig:Amplifier_illustration}
\end{figure*}

Here we present the first demonstration of a protocol, based on reversible squeezing, for ideally noiseless amplification of mechanical oscillator displacements.  This amplification method is applicable to any harmonic oscillator where reversible squeezing can be implemented faster than system decoherence. Our protocol is shown in Fig.~\ref{fig:Amplifier_illustration}. By first squeezing the motional ground state, quantum fluctuations along a particular phase space quadrature are suppressed. A small initial displacement $\alpha_{i}$ (to be amplified) is then applied along the squeezed axis. At this stage, although the signal-to-noise ratio for measuring $\alpha_{i}$ has been improved by squeezing, resolution below the zero-point fluctuations would require a detection method with yet lower noise. Finally, by reversing the squeezing interaction, the oscillator returns to a minimum-uncertainty coherent state with a larger amplitude $\alpha_{f}=G\alpha_{i}$, where $G$ is the gain. Ideally, this process adds no noise in either quadrature.  For an oscillator described using creation and annihilation operators $\hat{a}^{\dagger}$ and $\hat{a}$, the amplification is given by the identity~\cite{Nieto1997}

\begin{equation}
\hat{D}(\alpha_{f})=\hat{S}^{\dagger}(\xi)\hat{D}(\alpha_{i})\hat{S}(\xi),
\end{equation}

\noindent where $\hat{D}(\alpha)=\exp(\alpha \hat{a}^{\dagger}-\alpha^{*} \hat{a})$ is the displacement operator, and $\hat{S}(\xi)=\exp(\frac{1}{2}(\xi^{*}\hat{a}^{2}-\xi\hat{a}^{\dagger2}))$ is the squeezing operator with the complex squeezing parameter  $\xi(r,\theta)=re^{i\theta}$. For arbitrary orientations of the displacement $\alpha_{i}$ with respect to the initial squeezing axis, $\alpha_{f}=\alpha_{i} \cosh(r)+\alpha^{*}_{i}e^{i\theta}\sinh(r)$. Maximum amplification is achieved if the displacement is along the squeezed axis, giving $G = e^{r}$. 

We demonstrate this technique using a single trapped $^{25}$Mg$^{+}$ ion as the mechanical oscillator~\cite{Sup}.  The ion is held ${\simeq}\,30\,\mu$m above a linear surface-electrode radio-frequency trap~\cite{Seidelin2006, Ospelkaus2011}, which is cryogenically cooled to 18\,K.  Experiments are performed on a radial motional mode of the ion with frequency $\omega_{r} \simeq 2\pi\times 6.3\,$MHz, energy eigenstates denoted by $\ket{n}$, and zero-point wavefunction extent of $\simeq5.7\,$nm \cite{Sup}. To analyze the motional state, we use qubit states $\ket{\downarrow}\equiv\ket{F=3, m_F=1}$ and $\ket{\uparrow}\equiv\ket{F=2, m_F=1}$ within the $^{2}S_{1/2}$ electronic ground state hyperfine manifold, where $F$ is the total angular momentum and $m_F$ is its projection along the direction of the quantization magnetic field of approximately 21.3$\,$mT. The qubit transition frequency $\omega_{0}\simeq 2\pi\times 1.686\,$GHz is first-order insensitive to magnetic field fluctuations, giving a qubit coherence time longer than 200\,ms~\cite{Ospelkaus2011}. The qubit state can be manipulated with resonant microwave carrier pulses. In each experiment, the ion is initialized in the electronic and motional ground state $\ket{\downarrow}\ket{0}$ with optical pumping, resolved-sideband laser cooling \cite{Monroe1995}, and microwave pulses. Qubit readout is accomplished by applying a laser resonant with the $^{2}S_{1/2}\,\,$$\leftrightarrow$$\,\, ^{2}P_{3/2}$ cycling transition and detecting state-dependent ion fluorescence. We analyze the motional state of the ion by applying sideband interactions to map it onto the qubit states~\cite{Meekhof1996, Kienzler2015}. Applying a sideband interaction for various durations results in qubit Rabi oscillations with multiple frequency components whose amplitudes depend on the Fock state populations. We generate these interactions using oscillating magnetic field gradients~\cite{Ospelkaus2008, Ospelkaus2011}.  The blue sideband (BSB) interaction induces transitions between the states $\ket{\downarrow}\ket{n}$ and $\ket{\uparrow}\ket{n+1}$ with Rabi frequencies proportional to $\sqrt{n+1}$. The red sideband (RSB) interaction drives transitions between $\ket{\downarrow}\ket{n}$ and $\ket{\uparrow}\ket{n-1}$ with Rabi frequencies proportional to $\sqrt{n}$, and will not cause a qubit transition if the ion is in $\ket{\downarrow}\ket{0}$.  

\begin{figure*}
	\centering
	\includegraphics[scale=1.0]{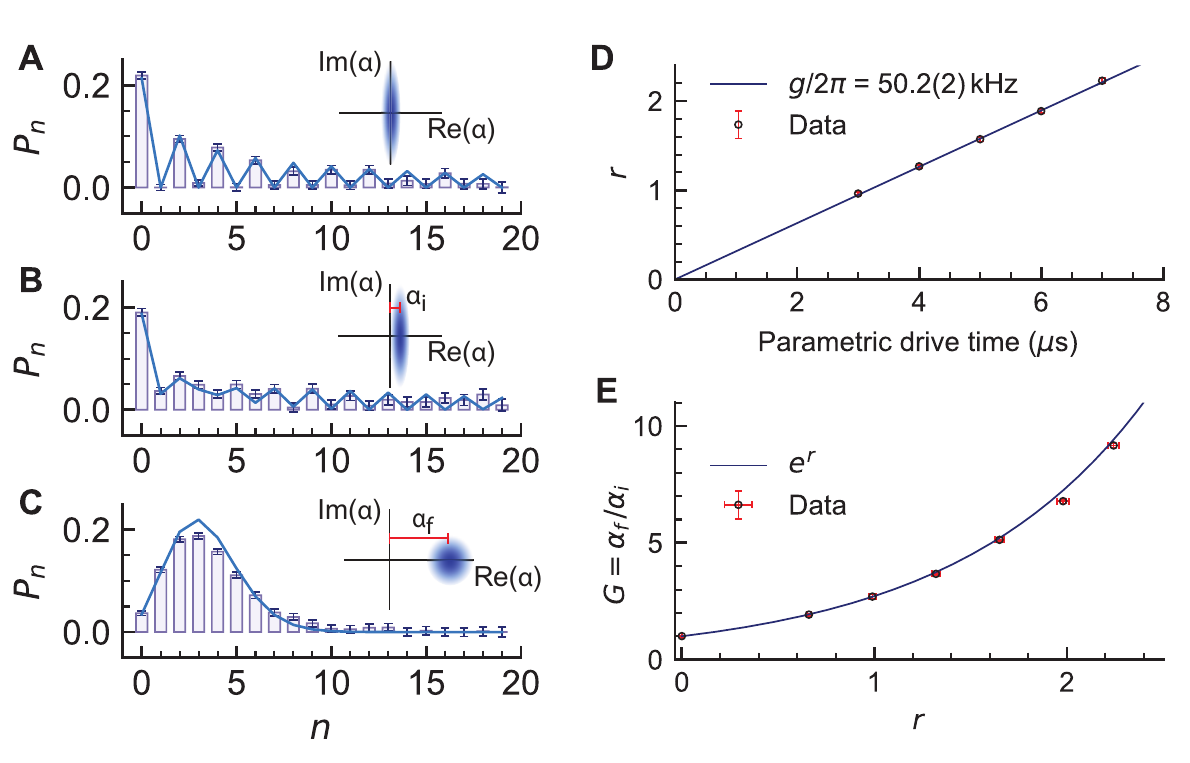}
	\caption{\textbf{Fock state population analysis.} A, B, and C show histograms of Fock state populations extracted by fitting to BSB Rabi oscillations. Vertical bars are derived by fitting to an unconstrained population distribution.  Solid blue lines are fits assuming parameterized functional forms of the ideal Fock state populations, yielding values of $r$, $\alpha_i$, and $\alpha_f$~\cite{Sup}.  Insets show Wigner function illustrations of the corresponding motional states. (\textbf{A}) Initial squeezed motional ground state with $r=2.26(2)$. (\textbf{B}) After displacing this state by $\alpha_{i} =0.200(2)$. (\textbf{C}) Final coherent state with amplitude $\alpha_{f}=1.83(1)$, following the reversed squeezing operation. The initial displacement is amplified by ${G=\alpha_{f}/\alpha_{i}=9.17(9)}$. (\textbf{D}) Squeezing parameter $r$ (black circles) as a function of the parametric drive duration. The solid line is a linear fit whose slope gives the parametric coupling strength $g$. (\textbf{E}) Measured gain (black circles) as a function of the squeezing parameter $r$. The solid line is the theoretical gain $G=e^{r}$. Error bars in all figures indicate one standard deviation of the mean.}
    \label{fig:para_amp}
\end{figure*}

Squeezing of the motional state is accomplished by applying an oscillating potential at twice the motional frequency ($2\omega_{r}$) to the radio-frequency electrodes of the trap~\cite{Heinzen1990}. This modulates the confining potential for the ion, yielding the interaction picture Hamiltonian~\cite{Sup}

\begin{equation}
\hat{H}=i\hbar \frac{g}{2}(\hat{a}^{2}e^{-i\theta}-\hat{a}^{\dagger2}e^{i\theta}),
\end{equation}

\noindent where $g$ is the parametric coupling strength and $\theta$ is the phase of the parametric modulation. Applying this Hamiltonian for duration $t$ implements the unitary squeezing operator $\hat{S}(\xi)$ with $r=gt$. While electronic parametric modulation has been used with single ions to squeeze a thermal state of motion~\cite{Natarajan1995} and for phase-sensitive parametric amplification of highly displaced thermal states~\cite{Yu1993}, it has not previously been implemented on pure quantum states. Optical forces can also be used for parametric modulation~\cite{Meekhof1996, Maslennikov2017}, but decoherence due to photon scattering and higher order nonlinearities in the optical field have limited the achievable squeezing~\cite{Meekhof1996}. Squeezed mechanical oscillator states can also be prepared using dissipative reservoir engineering~\cite{Kienzler2015, Wollman2015}. However, this is not a unitary squeezing operation, as is required for the amplification method described here. 

We characterize our squeezing process using motional sideband analysis to extract Fock state populations~\cite{Sup} as shown in Fig.~\ref{fig:para_amp}.  To characterize the unitarity of our squeezing operations, we measure the ground state population after squeezing and anti-squeezing, $\bra{0}\hat{S}^\dagger\hat{S}\ket{0}$.  For ${r<2}$, this population is ${\approx 0.98}$, which is consistent with the measured value without squeezing and anti-squeezing (${r=0}$). The population in ${n=0}$ remains above 0.93 for ${r<2.37(3)}$, or {$20.6(3)\,$dB} of squeezing~\cite{Sup}.  The calibrated parametric coupling strength is ${g=2\pi\times50.2(2)\,}$kHz, equivalent to a squeezing rate of 2.75(2)\,dB/$\mu$s. 

We use this unitary squeezing interaction to demonstrate amplification of harmonic oscillator displacements (see Fig.~\ref{fig:Amplifier_illustration}). Displacements are implemented by applying an oscillating potential resonant with the motional mode (at frequency $\omega_{r}$) to an electrode of the ion trap~\cite{Sup, Wineland1998}. All control fields are digitally synthesized with the same reference clock, enabling stable and deterministic control of the relative phases between the displacement, squeezing, sideband, and carrier interactions. At each stage of the amplification process, we verify the Fock state composition of the ion's motional state using sideband analysis with example data shown in Figs.~\ref{fig:para_amp}A-C. As shown in Fig.~\ref{fig:para_amp}E, the measured gain for various values of the squeezing parameter $r=gt$ closely follows the theoretically expected exponential growth of the coherent state amplitude. 

\begin{figure*}
	\centering
	\includegraphics[scale=1.0]{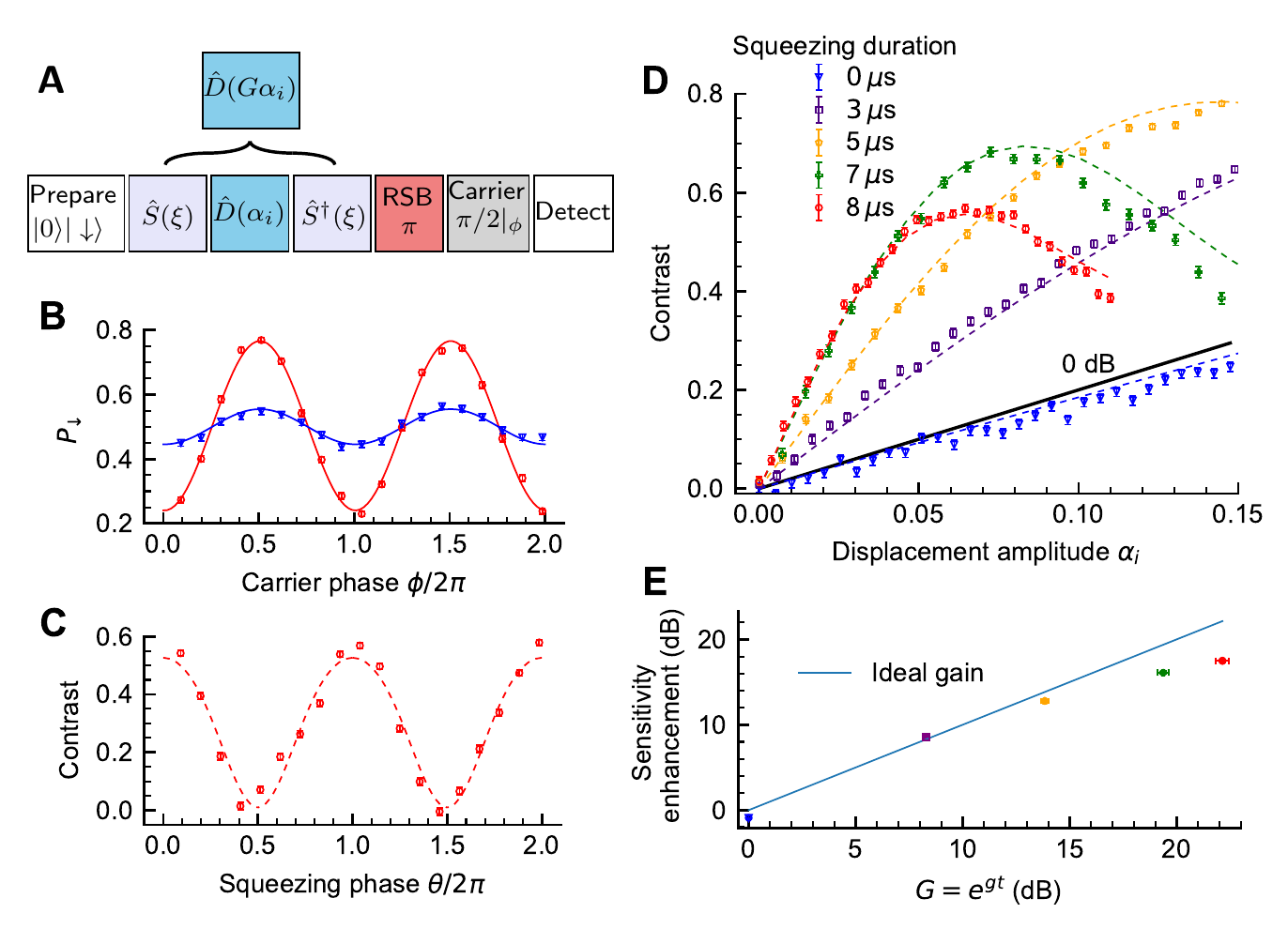}
	\caption{Measurement sensitivity enhancement. (A) Pulse sequence for displacement sensing protocol with phase-sensitive red-sideband (PSRSB) detection. (B) Population in $\ket{\downarrow}$ as a function of the carrier $\pi/2$ pulse phase. Blue diamonds are data with no squeezing and red circles are data with amplification. Solid lines show sinusoidal fits to the data. (C) Contrast of the carrier phase scan, as shown in (B), as a function of the squeezing phase $\theta$ for a fixed displacement. (D) Contrast as a function of  the displacement amplitude $|\alpha_i|$ for different initial squeezing pulse durations. Each data point is calculated from $\approx 10^{4}$ experiments.  The data shown in (B) and (C) have initial $|\alpha_i|=0.0550(6)$ and a squeezing duration of $t=8\,\mu$s (nominally $r=2.54(3)$). The solid black line is the maximum theoretical contrast without squeezing. Dashed lines in (C) and (D) are derived from a numerical model that includes motional decoherence. (E) Measurement sensitivity enhancement in the linear small-displacement regime as a function of the ideal gain $G=e^{gt}$. For each squeezing duration, the enhancement is measured by dividing the slope of the contrast for $C\lsim 0.25$ (obtained by fitting a straight line to data points in (D) with $C\lsim 0.25$) by the slope of the $0\,$dB black line. All error bars represent one standard deviation of the mean.}
    \label{fig:PSRS}
\end{figure*}

Using this amplification technique, we achieve increased sensitivity when measuring displacements much smaller than the zero-point fluctuations. To map the final displacements $\alpha_f$ onto the qubit states, we use a phase-sensitive red sideband (PSRSB) method~\cite{Hempel2013} shown in Fig.~\ref{fig:PSRS}A.  In this method, a displacement of the motional ground state results in a probability of measuring $\ket{\downarrow}$ of $P_{\downarrow} = \textstyle{\frac{1}{2}(1-C(|\alpha_f|)\cos\phi)}$, where $C(|\alpha_f|)$ is the signal contrast and $\phi$ is the phase of a carrier $\pi/2$ pulse, which follows the red sideband $\pi$ mapping pulse. For $|\alpha_f|\ll1$, $C(\alpha_f)\approx 2|\alpha_f|$. In comparison, simply measuring the qubit directly after the RSB $\pi$-pulse gives a signal $P_{\downarrow}\propto|\alpha_f|^{2}$.  Without amplification, $\alpha_f=\alpha_i$ and the PSRSB contrast is $C(|\alpha_i|)$.  With amplification, the initial displacement amplitude $\alpha_i$ is ideally increased by a factor of $G$ and the PSRSB contrast becomes $C(|G\alpha_i|)$.  This increase in contrast is shown in Fig.~\ref{fig:PSRS}B, where the presence of oscillations for the state after amplification indicates that it has a well-defined motional phase.  The carrier phase dependence in this figure is a feature of the PSRSB method, not of the amplification protocol.  The phase-sensitive nature of the amplification protocol is shown in Fig.~\ref{fig:PSRS}C, where the contrast $C$ of the PSRSB fringe is plotted against the squeezing phase $\theta$ for a fixed displacement. Maximum amplification is achieved when the displacement is oriented along the squeezed axis of the initial squeezed state in motional phase space (see Fig.~\ref{fig:Amplifier_illustration}). Figure~\ref{fig:PSRS}D shows the measured signal contrast as a function of $|\alpha_i|$ for various parametric drive durations. For each displacement, the contrast is defined as $C\equiv P_{\downarrow, max}-P_{\downarrow,min}$, where $P_{\downarrow, max}$ and $P_{\downarrow,min}$ are the maximum and minimum, respectively, of the fringes shown in Fig.~\ref{fig:PSRS}B. The uncertainty in measuring the contrast is $\sigma(C)=\sqrt{\sigma(P_{\downarrow, max})^{2}+\sigma(P_{\downarrow,min})^{2}}$, where $\sigma(P_{\downarrow,max(min)})^{2}$ is the variance of the projection noise associated with measuring $P_{\downarrow, max(min)}$. The signal-to-noise ratio (SNR) for a displacement measurement is then $s(G)=C(G\alpha_i)/\sigma(C(G\alpha_i))$. For a given number of experiments, amplification allows the SNR for a displacement measurement to be improved in comparison to the ideal PSRSB measurement with no squeezing (black solid line in Fig.~\ref{fig:PSRS}D), giving a measurement sensitivity enhancement of $s(G)/s(G=1)$. For measurements where $C\lsim 0.25$, the contrast varies linearly with $|\alpha_i|$, and the gain in contrast $C(G\alpha_i)/C(\alpha_i)$ sets a lower bound (which becomes exact as $|\alpha_i|\rightarrow0$) on the measurement sensitivity enhancement, because the projection noise decreases monotonically with increasing contrast. Increasing the squeezing results in increased contrast for $|\alpha_f|\ll1$, up to a squeezing time of approximately 8$\,\mu$s (corresponding to $r=2.54(3)$, and ideally 22.0(3)$\,$dB of squeezing). Here we achieve a contrast gain of $7.5(3)$, corresponding to $17.5(3)\,$dB enhancement in measurement sensitivity, which gives a ${\sim 56\times}$ reduction in the number of measurements required to achieve a given SNR.  For larger squeezing durations, degradation of the contrast due to background motional heating and dephasing in our trap prevent further increase in gain. This is not a limitation of the amplification process or our squeezing method. We note that with amplification we can achieve an SNR of 1 for measuring a displacement of one Bohr radius ($\simeq0.0529 \,$nm, corresponding to $\alpha= 0.00465$), $108$ times below the extent of the ground-state vacuum fluctuations ($\alpha = 0.5$), in $\sim200$ experiments.

In conclusion, we have implemented a fast unitary squeezing interaction in a simple mechanical oscillator and used it to amplify and detect coherent motional displacements that are significantly smaller than the quantum zero-point fluctuations. This amplification technique can enhance measurement sensitivity in protocols that use phase-stable displacements, such as photon-recoil spectroscopy~\cite{Hempel2013, Wan2014}, where the phase of momentum kicks from photon absorption can be controlled by modulating the photon source. The parametric modulation used for squeezing can also be combined with a spin-dependent force to enhance phonon-mediated spin-spin interactions~\cite{Ge2018}, which are used to create entanglement in quantum simulation and quantum information processing experiments~\cite{Pezze2018, Haffner2008}.  Our methods are also applicable to the generation of exotic non-classical motional states~\cite{Lo2015} and to continuous-variable quantum information processing~\cite{Fluhmann2018}. Finally, we note that the squeezing, displacement, spin-motion coupling, and qubit control interactions employed in this work are all generated without lasers, thereby eliminating spontaneous emission, simplifying control of relative phases, and enabling use with other charged particles lacking optical transitions such as electrons, positrons, and (anti-)protons~\cite{Heinzen1990,Wineland1998}.  

\bibliography{scibib}
\bibliographystyle{Science}

\section*{Acknowledgments}
We thank W. Ge, D. Kienzler, and D. M. Lucas for stimulating discussions, and S. M. Brewer and S. S. Kotler for comments on the manuscript. These experiments were performed using the ARTIQ control system. S.C.B., R.S., and D.T.C.A. are Associates in the Professional Research Experience Program (PREP) operated jointly by NIST and the University of Colorado Boulder under award 70NANB18H006 from the U.S. Department of Commerce, National Institute of Standards and Technology. This work was supported by ARO, ONR, and the NIST Quantum Information Program. This paper is a contribution of NIST and is not subject to US copyright.

\section*{Supplementary materials}
Supplementary Text\\
Figs. S1 to S3\\
Reference \textit{(41-43)}\\

\renewcommand{\thefigure}{S\arabic{figure}}
\setcounter{figure}{0}
\setcounter{equation}{0}
\renewcommand{\theequation}{S\arabic{equation}}

\section{Introduction}

This supplementary information contains further details about our experimental apparatus and data analysis procedures. Details concerning the experimental implementation of the parametric modulation Hamiltonian are given in Section~2. A discussion of motional-sideband analysis and characterization of squeezing interactions is presented in Section~3. The phase-sensitive red sideband (PSRSB) method is described in Section~4.      

\section{Parametric drive}

Consider a harmonically confined one-dimensional mechanical oscillator with mass $m$ and frequency $\omega_{r}$. If the potential of the oscillator is modulated at a frequency $\omega_{p}$, the Hamiltonian can be written as

\begin{equation}
\hat{H}=\frac{\hat{p}^{2}}{2m}+\frac{1}{2}m\omega_{r}^{2}\hat{x}^{2}-\hbar g\sin(\omega_{p}t-\theta)\left(\frac{\hat{x}}{x_{0}}\right)^{2},
\end{equation}

\noindent where $g$ is the modulation strength, $\theta$ is the phase of the modulation drive, $\hat{x}$ is the position operator, and  $\hat{p}$ is the momentum operator. In terms of the creation and annihilation operators $\hat{x}=x_{0}(\hat{a}+\hat{a}^{\dagger})$ and $\hat{p}=ip_{0}(\hat{a}^{\dagger}-\hat{a})$, where $x_{0}=\sqrt{\textstyle{\frac{\hbar}{2 m \omega_{r}}}}$ is the ground state wavefunction extent and $p_{0}=\hbar/(2x_{0})$, the Hamiltonian becomes

\begin{equation}
\hat{H}=\hbar\omega_{r}\left(\hat{a}^{\dagger}\hat{a}+\frac{1}{2}\right)-\hbar g\sin(\omega_{p}t-\theta)(\hat{a}^{\dagger 2}+\hat{a}^{2}+2\hat{a}^{\dagger}\hat{a}+1).
\end{equation}

\noindent Transforming into the interaction picture with respect to the bare $(g=0)$ harmonic oscillator Hamiltonian gives the interaction Hamiltonian

\begin{equation}
\hat{H}_{I}=-\hbar g\sin(\omega_{p}t-\theta)(\hat{a}^{\dagger 2}e^{2i\omega_{r}t}+\hat{a}^{2}e^{-2i\omega_{r}t}+2\hat{a}^{\dagger}\hat{a}+1).
\end{equation}

\noindent If we make the rotating wave approximation by neglecting terms oscillating at $\pm(2\omega_{r}+\omega_{p})$ and $\omega_{p}$, $\hat{H}_{I}$ reduces to

\begin{equation}
\hat{H}_{I}=-\hbar \frac{g}{2i}\left(\hat{a}^{2}e^{-2i\omega_{r}t+i\omega_{p}t-i\theta}-\hat{a}^{\dagger 2}e^{2i\omega_{r}t-i\omega_{p}t+i\theta}\right).
\end{equation}

\noindent If $\omega_{p}=2\omega_{r}$,

\begin{equation}
\hat{H_{I}}=i\hbar\frac{ g}{2}(\hat{a}^{2}e^{-i\theta}-\hat{a}^{\dagger2}e^{+i\theta}),
\end{equation}

\noindent which is the Hamiltonian describing degenerate parametric amplification of a single-mode boson field~\cite{Walls1994}.\\

\begin{figure*}
	\centering
	\includegraphics[scale=1.0]{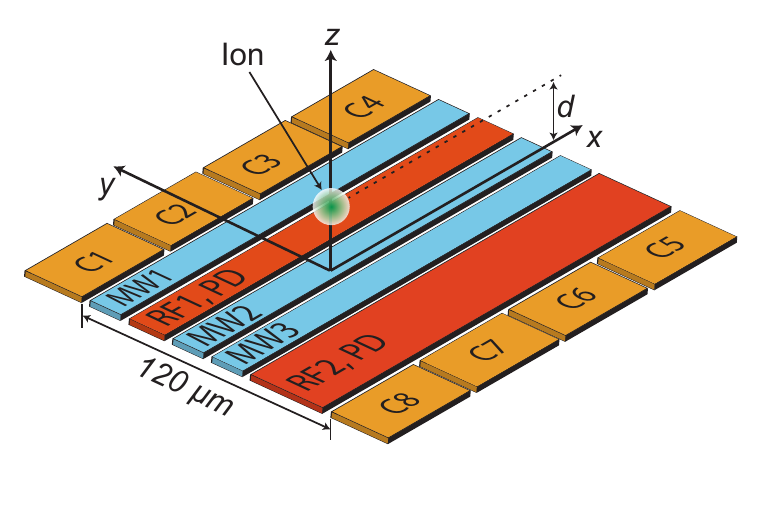}
    \caption{Schematic diagram of the relevant portion of the surface-electrode ion trap. The ion is trapped at a distance $d=30\,\mu$m above the surface. Static potentials applied to electrodes C1-C8, along with an rf potential applied to RF1 and RF2, provide confinement for the ion in three dimensions. For radial oscillator modes in the $y-z$ plane, a unitary squeezing interaction is achieved by applying a parametric drive (PD) directly to the radio-frequency electrodes. Electrodes C1-C8 provide confinement along the axial ($x$) direction. Displacements are implemented by applying an oscillating potential to C4. Electrodes MW1-MW3 carry oscillating currents that generate magnetic fields and field gradients for qubit control and spin-motion coupling.}
    \label{fig:Trap}
\end{figure*}

In this work, the mechanical oscillator is the harmonic motion of a single atomic ion in a radio-frequency Paul trap. Figure~\ref{fig:Trap} shows a schematic diagram of our surface-electrode ion trap. Confinement in the $y-z$ radial plane normal to the trap axis is achieved by applying an rf potential with ${\simeq 30\,}$V amplitude at ${\simeq 68\,}$MHz to electrodes RF1 and RF2.  The radial normal mode frequencies are approximately 6.3\,MHz and 7.5\,MHz. We use the lower frequency radial mode whose oscillation direction is close to parallel to the trap surface.  Parametric modulation of this mode is implemented by applying an oscillating potential at twice the motional frequency (${\simeq12.6\,}$MHz) to the rf electrodes of the trap. From our measured parametric coupling strength of ${g=50.2(2)\,}$kHz, we use an electric field simulation of the trap to estimate a parametric drive amplitude of ${\sim 20\,}$mV. Applying the parametric drive to the rf electrodes ensures that the electric field curvature at the position of the ion can be strongly modulated with minimal residual electric field oscillating at the parametric drive frequency.  

\section{Motional-state sideband analysis and squeezed state characterization}

As described in~\cite{Meekhof1996, Wineland1998}, the Fock state populations of the motional state of a trapped ion can be analyzed by applying a motional sideband pulse for varying interaction times.  The blue sideband (BSB) is described by the Hamiltonian ${\hat{H}_{BSB}=\frac{\hbar\Omega}{2}(\hat{\sigma}_{+}\hat{a}^{\dagger}+\hat{\sigma}_{-}\hat{a})}$, where $\Omega$ is the sideband Rabi frequency and $\hat{\sigma}_{+} =\ket{\uparrow}\bra{\downarrow}$ and $\hat{\sigma}_{-} =\ket{\downarrow}\bra{\uparrow}$ are spin-flip operators.  The red sideband (RSB) is described by ${\hat{H}_{RSB}=\frac{\hbar\Omega}{2}(\hat{\sigma}_{+}\hat{a}+\hat{\sigma}_{-}\hat{a}^{\dagger})}$.  The sideband pulses map the motional state information onto the internal qubit states. For the blue sideband, the probability of measuring the $\ket{\downarrow}$ is given by

\begin{equation}
P_{\downarrow}(t)=\frac{1}{2}(1+\sum_{n=0}^{\infty}P_{n}e^{-\gamma_{n}t}\cos(\Omega_{n+1,n}t)).
\label{eq:Pop}
\end{equation}

In this work we use a microwave gradient sideband~\cite{Ospelkaus2011} with a sideband Rabi frequency of ${\Omega \simeq 2\pi\times 1.1\,}$\,kHz. Using this sideband, ${\Omega_{n+1,n}=\Omega\sqrt{n+1}}$ to a good approximation even for large $n$ (experimentally verified for $n$ up to 50). The $\gamma_{n}$ are phenomenological decay constants assumed to obey ${\gamma_{n}=\gamma\sqrt{n+1}}$. For a particular motional state, the probability of the {$n$\textsuperscript{th}} Fock state being occupied  is $P_{n}$. For coherent states, the Fock state decomposition is

\begin{equation}
\ket{\alpha}=e^{-|\alpha|^2/2}\sum_{n=0}^{\infty}\frac{\alpha^{n}}{\sqrt{n!}}\ket{n},
\end{equation}

\noindent and

\begin{equation}
P_{n}=|\braket{n|\alpha}|^{2}=\frac{e^{-|\alpha|^{2}}|\alpha|^{2n}}{n!}.
\label{eq:coherent}
\end{equation}

\noindent For a squeezed state characterized by the complex squeezing parameter $\xi=re^{i\theta}$, 

\begin{equation}
\ket{\xi}=\frac{1}{\sqrt{\cosh r}}\sum_{n=0}^{\infty}\frac{e^{in\theta}(-\tanh r)^{n}\sqrt{(2n)!}}{2^{n}n!}\ket{2 n},
\label{eq:squeezed_state_n}
\end{equation}

\noindent showing that the odd Fock state amplitudes are zero. The even populations are given by 

\begin{equation}
P_{2n}=\frac{(\tanh r)^{2n}}{\cosh r}\frac{(2n)!}{(2^{n}n!)^{2}}.
\label{eq:squeezed_state_Pn}
\end{equation}

\begin{samepage}
\noindent For a state that is first squeezed with squeezing parameter $\xi$ and then displaced by $\alpha$ \cite{Gerry2005}
\begin{eqnarray}
\label{eq:disp_squeezed_Pn}
P_{n}&=&|\braket{n|\alpha,\xi}|^{2}=\frac{(\frac{1}{2}\tanh r)^{n}}{n! \cosh r}\left|H_{n}\left(\frac{\alpha \cosh r+\alpha^{*}e^{i\theta}\sinh r}{\sqrt{e^{i\theta}\sinh(2r)}} \right)  \right|^{2}\nonumber\\
&&\hspace{2.5cm}\times\exp\left[-|\alpha|^{2}-\frac{1}{2}(\alpha^{*2}e^{i\theta}+\alpha^{2}e^{-i\theta})\tanh r\right], 
\end{eqnarray}
\end{samepage}
\noindent where $H_n$ are Hermite polynomials.  

\begin{figure*}
	\centering
	\includegraphics[scale=1.0]{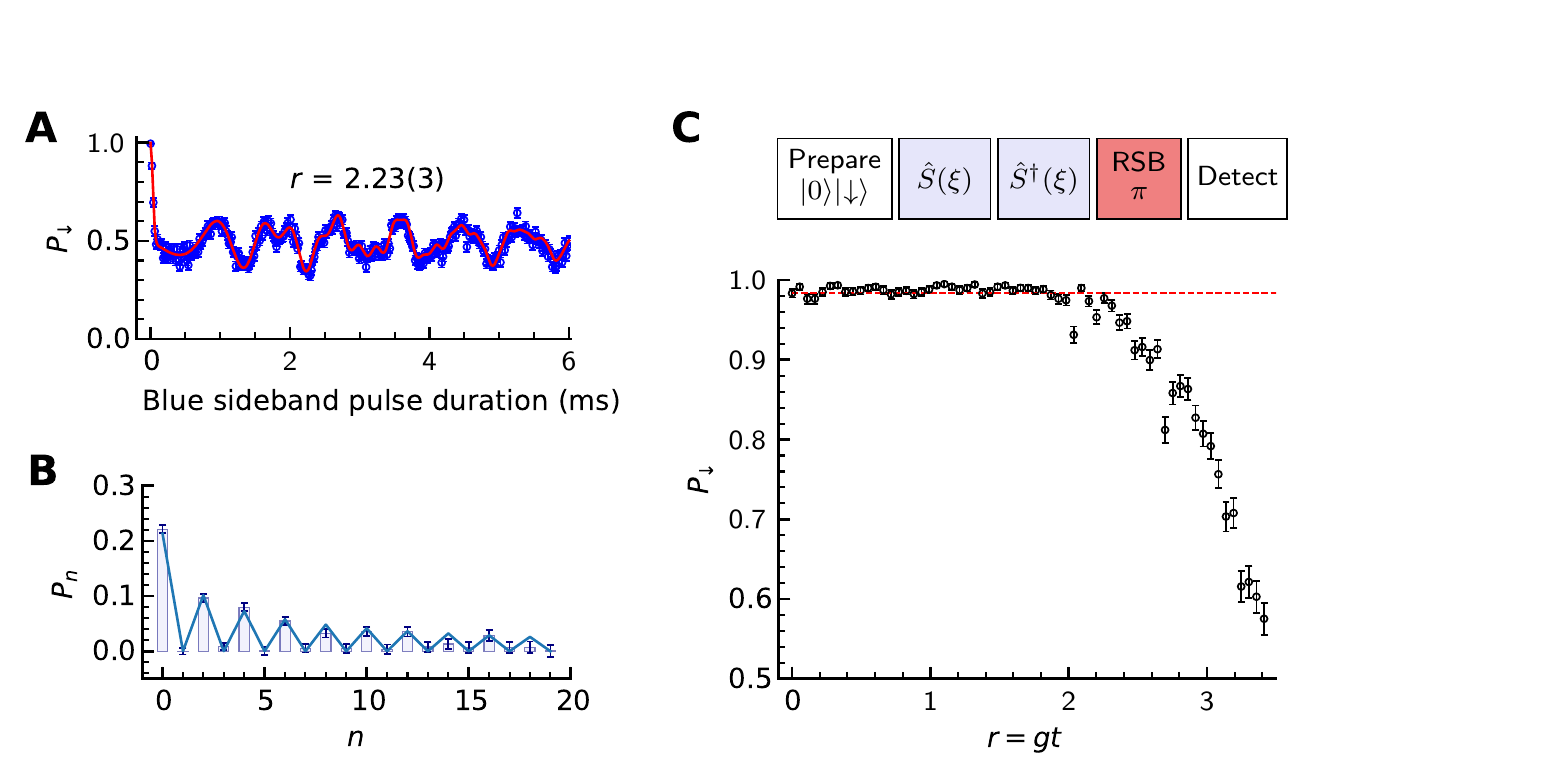}
	\caption{Characterization of the squeezing operation. (A) Probability of measuring $\ket{\downarrow}$ for a squeezed state as a function of the blue sideband analysis pulse duration. The solid red curve is a fit of Eq.~\ref{eq:Pop} to the data, assuming a squeezed state distribution (Eq.~\ref{eq:squeezed_state_Pn}), giving a squeezing parameter of $r=2.23(3)$. (B) Fock state populations obtained by fitting Eq.~\ref{eq:Pop} to the experimental data without assuming a distribution for $P_{n}$. The solid blue curve indicates the populations of an ideal squeezed state with $r=2.23$. (C) Probability of measuring $\ket{\downarrow}$ after a red sideband (RSB) analysis $\pi$ pulse that measures population not returned to the motional ground state after reversing the squeezing interaction for various parametric drive durations.  Even without squeezing, there is $\sim\!2\,\%$ residual population in $\ket{\uparrow}$ after the RSB pulse due to imperfect state preparation and motional heating during the analysis pulse. This background is indicated by the dashed red line.}
    \label{fig:squeezed_states}
\end{figure*}

Results such as those shown in Fig.~\ref{fig:para_amp} are obtained as follows. The ion is first prepared in the motional and electronic ground state $\ket{0}\ket{\downarrow}$. Interactions that transform the motional state are then applied as described in the main text.  A blue sideband pulse is then applied for a variable duration and the internal state is detected. Fitting Eq.~\ref{eq:Pop} to the data as a function of the interaction duration, and assuming an underlying distribution for $P_{n}$ from Eq.~\ref{eq:coherent}, \ref{eq:squeezed_state_Pn}, or \ref{eq:disp_squeezed_Pn} as appropriate, allows us to extract best fit values for parameters characterizing the assumed motional state. We also separately fit Eq.~\ref{eq:Pop} to the data assuming no particular underlying distribution for $P_{n}$ giving the Fock state populations of the motional state being analyzed. Error bars for fitting parameters are calculated as the square root of the corresponding diagonal elements of the least-squares covariance matrix. Example data for a squeezed state are shown in Fig.~\ref{fig:squeezed_states}A. 

Although BSB analysis enables the determination of Fock state populations, it does not provide information about the phase coherence between the Fock states. To investigate the phase coherence, the squeezing operation is applied and then reversed. An ideal squeezing interaction is unitary so reversing the squeezing operation should bring the motional state back to the ground state i.e. ${\hat{S}^{\dagger}(\xi)\hat{S}(\xi)\ket{0}=\ket{0}}$. Population not in the motional ground state is mapped onto $\ket{\uparrow}$ by an RSB $\pi$ pulse. Therefore imperfections in the squeezing process that result in population not returning to the motional ground state reduce the probability of measuring $P_\downarrow$. As shown in Fig.~\ref{fig:squeezed_states}B, the population $P_{\downarrow}$ does not change appreciably until $r\simeq2.0\,$. We also use BSB analysis to measure the population remaining in the ground state after reversing the squeezing interaction. For $t\,=\,7.5\,\mu$s, corresponding to $r\,=\,2.37(3)$ and ideally 20.6(3)$\,$dB of squeezing, $93.8(7)\,\%$ of the population has returned to the ground state. Increasing loss in coherence for longer parametric drive durations is attributed to imperfections in the parametric drive electronics, motional decoherence, and trap anharmonicities.       

\section{Measurement of weak displacements}

We use the phase-sensitive red sideband method~\cite{Hempel2013} to measure displacements of the motional ground state. For displacements with $|\alpha|\ll1$, this method gives measurement sensitivities proportional to $|\alpha|$. This linear scaling enables straightforward measurement of the amplification gain and is desirable for measuring small displacements. We first give a intuitive description that is valid for small displacements. The ion is first prepared in the motional ground state $\ket{\psi}=\ket{\downarrow}\ket{0}$. A small displacement of the motional ground state ($|\alpha|\ll1$) results in a small population transfer out of the ground state, predominantly to the $n=1$ Fock state. In the interaction picture with respect to the qubit and the harmonic oscillator~\cite{Wineland1998}:

\begin{eqnarray}
\ket{\psi}&\rightarrow&\hat{D}(\alpha)\ket{0}\ket{\downarrow}\\
      &=&(\hat{I}+\alpha\hat{a}^{\dagger}-\alpha^{*}\hat{a}+...)\ket{0}\ket{\downarrow}\\
      &\approx&(\ket{0}+\alpha\ket{1})\ket{\downarrow}.
\end{eqnarray}

\begin{figure*}
	\centering
	\includegraphics[scale=1.0]{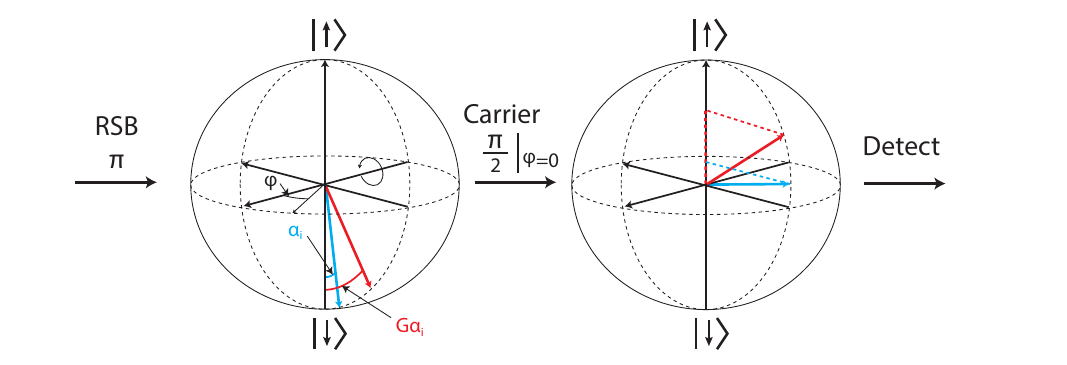}
	\caption{Bloch sphere illustration of the phase-sensitive red sideband readout scheme.}
    \label{fig:Bloch_spheres}
\end{figure*}

\noindent A red sideband $\pi$ pulse maps the the population in $\ket{1}\ket{\downarrow}$ onto $\ket{0}\ket{\uparrow}$, but leaves the population in $\ket{\downarrow}\ket{0}$ unchanged. The state becomes:

\begin{equation}
\ket{\psi} \approx (\ket{\downarrow}-|\alpha|\ket{\uparrow})\ket{0}
\end{equation}

\noindent for the case where the phase of the beatnote between the RSB pulse and the displacement has been set to give a qubit rotation about the $\phi=0$ axis on Bloch sphere shown in Fig~\ref{fig:Bloch_spheres}. The subsequent carrier $\pi/2$ rotation about the $\phi$ axis gives

\begin{equation}
\ket{\psi} \approx \frac{1}{\sqrt{2}}\left [(1-|\alpha| e^{+i\phi})\ket{\downarrow}-(|\alpha|+ e^{-i\phi})\ket{\uparrow}\right ]\ket{0},
\end{equation}

\noindent showing the state vector has a phase-sensitive projection onto the polar axis of the Bloch sphere. The probability of finding the system in $\ket{\downarrow}$ is then

\begin{samepage}
\begin{eqnarray}
P_{\downarrow}& \approx&\frac{1}{2}|1-|\alpha| e^{i\phi}|^{2} \nonumber\\
 & = &\frac{1+|\alpha|^{2}}{2}-|\alpha|\cos({\phi})  \nonumber\\
 & \approx &\frac{1}{2}-|\alpha|\cos({\phi}).
\end{eqnarray}
\end{samepage}

\noindent $P_{\downarrow}$ sinusoidally oscillates as a function of $\phi$ with contrast $2|\alpha|$. If the displacement is amplified, then the contrast becomes $2|G\alpha|$ with the requirement that $|G\alpha|\ll1$. Therefore $G$ can be extracted by measuring the slope of the contrast as a function of $|\alpha|$.

Next, we calculate $\ket{\psi}$ for all orders of $|\alpha|$. The state of the ion after the displacement, sideband interaction, and carrier pulse is 

\begin{equation}
\ket{\psi}=\hat{R}(\pi/2,\phi)\hat{U}_{RSB}\hat{D}(\alpha)\ket{\downarrow}\ket{0},
\end{equation}

\noindent where $\hat{R}(\pi/2,\phi)$ is a spin rotation matrix describing the $\pi/2$ carrier pulse and $\hat{U}_{RSB}$ is the resonant red sideband time evolution operator~\cite{Wineland1998}

\begin{equation}
\hat{U}_{RSB}=e^{-\frac{i\Omega t}{2}(\hat{\sigma}_{+}\hat{a}+\hat{\sigma}_{-}\hat{a}^{\dagger})},
\end{equation}

\noindent where $\Omega$ is the sideband Rabi frequency. The duration $t$ of the RSB interaction is set to give $\Omega t/2=\pi/2$. Measuring $\ket{\downarrow}$ gives

\begin{eqnarray}
\braket{\downarrow |\psi}&=&\bra{\downarrow}\hat{R}(\pi/2,\phi)\hat{U}_{RSB}\hat{D}(\alpha)\ket{\downarrow}\ket{0} \nonumber\\
&=&\frac{1}{\sqrt{2}}\left(\bra{\downarrow}+\bra{\uparrow}e^{i\phi}\right)\hat{U}_{RSB}\ket{\alpha}\ket{\downarrow}.
\end{eqnarray}

\noindent The probability of measuring $\ket{\downarrow}$ is therefore

\begin{eqnarray}
P_{\downarrow}&=&\frac{1}{2}\left[1+e^{i\phi}\bra{\downarrow}\bra{\alpha}\hat{U}_{RSB}^{\dagger}\ket{\downarrow}\bra{\uparrow}\hat{U}_{RSB}\ket{\alpha}\ket{\downarrow} \right. \nonumber\\
&&\left. +e^{-i\phi}\bra{\downarrow}\bra{\alpha}\hat{U}_{RSB}^{\dagger}\ket{\uparrow}\bra{\downarrow}\hat{U}_{RSB}\ket{\alpha}\ket{\downarrow}\right].
\label{eq:Pexact}
\end{eqnarray}

\noindent Evaluating the matrix elements gives

\begin{equation}
\bra{\uparrow}\hat{U}_{RSB}\ket{\alpha}\ket{\downarrow}=\sum_{l=0}^{\infty}\frac{(-i\frac{\pi}{2})^{2l+1}(1+\hat{n})^{l}}{(2l+1)!}\alpha\ket{\alpha},
\label{eq:evensum}
\end{equation}

\noindent and

\begin{equation}
\bra{\downarrow}\hat{U}_{RSB}\ket{\alpha}\ket{\downarrow}=\sum_{l=0}^{\infty}\frac{(i\frac{\pi}{2})^{2l}(\hat{n})^{l}}{(2l)!}\ket{\alpha},
\label{eq:oddsum}
\end{equation}

\noindent where $\hat{n}=\hat{a}^{\dagger}\hat{a}$ is the harmonic oscillator number operator. Setting the phase of the beatnote between the RSB pulse and the displacement so that the initial rotation is about the $\phi=0$ axis as before, and inserting equations~\ref{eq:evensum} and~\ref{eq:oddsum} into~\ref{eq:Pexact} gives

\begin{eqnarray}
P_{\downarrow}&=&\frac{1}{2}(1-|\alpha|f(|\alpha|)e^{i\phi}-|\alpha|f(|\alpha|)e^{-i\phi}) \nonumber\\
&=&\frac{1}{2}-|\alpha|f(|\alpha|)\cos(\phi),
\label{eq:Pexact_f}
\end{eqnarray}

\noindent where

\begin{equation}
f(|\alpha|)=\sum_{n=0}^{\infty}\frac{e^{-|\alpha|^{2}}|\alpha|^{2n}}{n!}\left(\frac{\cos(\frac{\pi}{2}\sqrt{n})\sin(\frac{\pi}{2}\sqrt{n+1})}{\sqrt{n+1}}\right).
\end{equation}

Equation~\ref{eq:Pexact_f} shows that $P_{\downarrow}$ oscillates sinusoidally with amplitude $|\alpha|f(|\alpha|)$ as a function of $\phi$. If squeezing is used to amplify the displacement, $|\alpha|\rightarrow G|\alpha|$. However, motional decoherence reduces the measured contrast. The theory curves shown in Fig.~\ref{fig:PSRS}B. in the main text are functions of the form $b+a\cos({\phi})$, with $b$ and $a$ as the fitting parameters. With $8\,\mu$s parametric drive duration for each squeezing pulse and $|\alpha|=0.055(2)$, the fitting parameters are $b=0.503(3)$ and $a=0.263(4)$. For the scan with no amplification, the fitting parameters are $b=0.500(2)$ and $a=0.055(3)$.

For various squeezing durations, we measure the contrast of the carrier phase scan signal (Fig.~\ref{fig:PSRS}B), as a function of the displacement amplitude, with the squeezing phase set to maximize the contrast. Experimental data and theory curves are shown in Fig.~\ref{fig:PSRS}D. The displacement amplitude is varied by changing the amplitude of the digital synthesizer that generates the resonant drive. The resonant drive pulse duration is fixed at 5$\,\mu$s. The reduction in contrast for larger squeezing durations is attributed mainly to motional heating and motional dephasing. The theory curves (dashed lines) shown in the Figs.~\ref{fig:PSRS}C and D are from numerical simulations incorporating these effects performed using QuTip~\cite{qutip2013}. The motional heating rate was independently measured to be $\dot{n}=20(3)$\,quanta\,s$^{-1}$. Motional heating is modelled by the Lindblad operators $\sqrt{\dot{n}}\hat{a}^{\dagger}$ and $\sqrt{\dot{n}}\hat{a}$. We model motional dephasing with the Lindblad operator $\sqrt{\Gamma}\,\hat{a}^{\dagger}\hat{a}$ with a dephasing rate $\Gamma = 18(6)\,$s$^{-1}$, obtained from a least-squares fit to the 8$\,\mu$s squeezing duration curve shown in Fig.~\ref{fig:PSRS}D.

\end{document}